\documentclass[sigconf]{acmart}

\usepackage{hyphenat}
\usepackage{enumitem}

\AtBeginDocument{%
  }
\setcopyright{acmlicensed}
\copyrightyear{2026}
\acmYear{2026}
\setcopyright{cc}
\setcctype{by-nc-nd}
\acmConference[MSR '26]{23rd International Conference on Mining Software Repositories}{April 13--14, 2026}{Rio de Janeiro, Brazil}
\acmBooktitle{23rd International Conference on Mining Software Repositories (MSR '26), April 13--14, 2026, Rio de Janeiro, Brazil}
\acmPrice{}
\acmDOI{10.1145/3793302.3793575}
\acmISBN{979-8-4007-2474-9/2026/04}

\begin{document}

\title{Why Are Agentic Pull Requests Merged or Rejected? \\An Empirical Study}
\author{Sien Reeve O. Peralta}
\author{Fumika Hoshi}
\author{Hironori Washizaki}
\author{Naoyasu Ubayashi}
\affiliation{%
  \institution{Waseda University}
  \city{Tokyo}
  \country{Japan}
}

\author{Inase Kondo}
\author{Yoshiki Higo}
\affiliation{%
  \institution{The University of Osaka}
  \city{Osaka}
  \country{Japan}
}

\author{Hiroki Mukai}
\author{Norihiro Yoshida}
\affiliation{%
  \institution{Ritsumeikan University}
  \city{Osaka}
  \country{Japan}
}

\author{Kazuki Kusama}
\affiliation{%
  \institution{Kyushu University}
  \city{Fukuoka}
  \country{Japan}
}

\author{Hidetake Tanaka}
\author{Youmei Fan}
\affiliation{%
  \institution{Nara Institute of Science and Technology}
  \city{Nara}
  \country{Japan}
}

\renewcommand{\shortauthors}{Peralta et al.}

\begin{abstract}
AI coding agents increasingly submit pull requests (Agentic-PRs) to open-source repositories, yet their performance is commonly assessed using merge and rejection outcomes alone. We hypothesized that these outcome labels do not reliably reflect agent capability without considering review interactions. To test this, we conducted a decision-oriented analysis of 11,048 closed Agentic Pull Requests, refined to 9,799 human-reviewed PRs, and manually inspected 717 representative cases to recover decision rationale from interaction artifacts. We found that rejection outcomes substantially overstate agent error: only 35.7\% of rejected PRs reflected clear agentic failures, while 31.2\% were driven by workflow constraints and 33.1\% lacked observable decision rationale. Among merged PRs, 15.4\% required explicit reviewer involvement through feedback or direct commits, and 5.5\% showed no visible interaction trace. We further observed systematic differences across agents, with Copilot and Devin more often embedded in reviewer-mediated workflows, while Codex and Cursor PRs were typically merged with minimal interaction. These results reject the assumption that PR outcomes alone capture agent performance and demonstrate the need for interaction-aware evaluation grounded in review behavior.
\end{abstract}



\maketitle

\section{Introduction}

AI coding agents increasingly submit \emph{Agentic Pull Requests (Agentic-PRs)} that participate directly in real-world software development. Evidence from the AIDev dataset shows that agent-authored PRs span thousands of repositories~\cite{li2025riseaiteammatessoftware}, reflecting a broader shift toward AI-native software engineering (SE~3.0)~\cite{hassan2024ainativesoftwareengineeringse}. Prior work reports productivity gains from AI-assisted coding~\cite{cui2025effects, peng2023impactaideveloperproductivity}, changes in review practices~\cite{alami2025humanmachinesoftwareengineers}, and growing concerns related to maintainability~\cite{oladele2025impact} and security risks~\cite{pearce2021asleepkeyboardassessingsecurity}.

Despite widespread adoption, it remains unclear why Agentic-PRs are merged or rejected. Existing studies predominantly rely on outcome-based measures such as merge rates or approval frequencies, which provide limited insight into how review decisions are formed. In practice, pull request decisions are shaped by interaction-level factors, including reviewer comments, CI outcomes, commit history, and workflow actions, rather than PR states alone.

This gap complicates the interpretation of both rejection and merge outcomes. Rejections may reflect workflow constraints or undocumented reviewer decisions rather than agentic failures, while merged PRs may depend on varying degrees of human involvement. Without examining interaction evidence, outcome-based evaluations risk misrepresenting agent capability.

To address this limitation, we analyze \emph{closed} Agentic-PRs with evidence of human evaluation. Our dataset includes 11,048 closed PRs from repositories with at least 500 stars, refined to 9,799 human-reviewed PRs. We manually inspect a stratified sample of 717 PRs to examine how merge and rejection decisions are formed in practice.

This study is guided by two research questions:
\begin{itemize}
    \item[RQ1:] What drives the rejection of Agentic-PRs? \\
    \emph{Motivation:} Rejections may stem from agentic failures, workflow constraints, or undocumented reviewer decisions. Distinguishing among these factors is necessary to avoid misattributing rejection outcomes to agent capability alone.

    \item[RQ2:] How does human involvement shape successful Agentic-PRs? \\
    \emph{Motivation:} We examine whether merged PRs reflect autonomous agent performance or depend on collaborative refinement through reviewer feedback or direct human intervention.
\end{itemize}

\textbf{Key Takeaway.}  
Analysis of 717 manually inspected Agentic-PRs demonstrates that pull request outcomes alone are insufficient for assessing agent performance. Only 35.7\% of rejections correspond to clear agentic failures, while 31.2\% are driven by workflow factors and 33.1\% lack observable decision rationale. Moreover, 15.4\% of merged PRs require explicit reviewer involvement, indicating that both success and failure outcomes depend on interaction-level processes that are invisible in merge or rejection labels alone.

\textbf{Contributions.}  
This paper makes three contributions:
(1) a decision-oriented empirical analysis that disentangles agentic failures from workflow-driven and undocumented rejection outcomes in Agentic-PRs;
(2) a quantitative characterization of human involvement in successful Agentic-PRs, including feedback loops and reviewer-applied commits; and
(3) empirical evidence that outcome-based metrics systematically conflate agent capability with repository workflows, motivating interaction-aware evaluation of AI coding agents.

\section{Motivation}

Prior empirical studies show that AI coding agents can improve developer productivity and accelerate task completion~\cite{cui2025effects, peng2023impactaideveloperproductivity}, while also introducing challenges related to review effort, maintainability, and security~\cite{oladele2025impact, pearce2021asleepkeyboardassessingsecurity}. Large-scale analyses of Agentic-PRs, including those enabled by the AIDev dataset~\cite{li2025riseaiteammatessoftware}, predominantly assess agent performance using outcome-based measures such as approval rates or merge frequency. However, these measures provide limited insight into how merge and rejection decisions are formed.

Software engineering research has shown that pull request decisions rely heavily on interaction-level signals, including discussion threads, CI outcomes, and iterative revisions, rather than final PR states alone~\cite{golzadeh2019effect}. Studies of AI-assisted code review similarly indicate that reviewer judgments are often conveyed through unstructured comments or small corrective actions that are not captured by metadata or formal review labels~\cite{cihan2024automatedcodereviewpractice, palvannan2023suggestionbotanalyzingimpact}. As a result, outcome-based evaluations may conflate agent behavior with workflow conventions, reviewer intervention, or project-specific practices.

A decision-oriented analysis grounded in interaction evidence is therefore necessary to distinguish agentic failures from workflow-driven and human-mediated outcomes. Such separation enables more accurate interpretation of agent performance and supports a clearer understanding of Human-AI collaboration in software development~\cite{watanabe2025useagenticcodingempirical, ming2025helpfulagentmeetsdeceptive, hou2024largelanguagemodelssoftware}.

\section{Data Collection}

\begin{figure*}[t]
  \centering
  \includegraphics[width=0.87\linewidth]{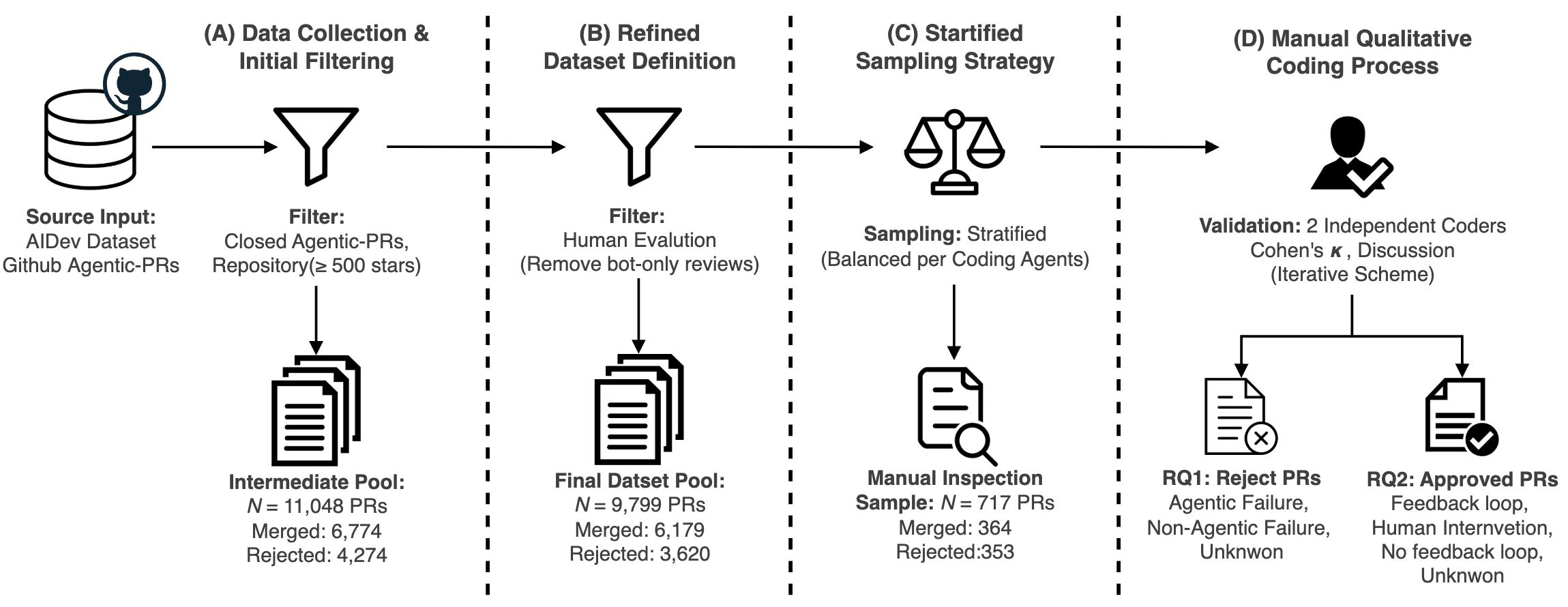}
  \caption{Overview of the study pipeline.}
  \Description{Overview of the study pipeline: (A) data collection and initial filtering, (B) refined dataset definition, (C) stratified sampling strategy, and (D) manual qualitative coding process.}
  \label{fig:flow}
\end{figure*}

Step A in Figure \ref{fig:flow} summarizes the initial construction of the dataset. We build on the AIDev dataset \cite{li2025riseaiteammatessoftware} and restrict the scope to closed Agentic-PRs, defined as pull requests with explicit merge or rejection decisions. To limit the analysis to repositories with established review activity, we apply a threshold of at least 500 stars. After this filtering step, the dataset contains 11,048 Agentic-PRs, including 6,774 merged and 4,274 rejected submissions.

The refined dataset definition is shown in step B in Figure \ref{fig:flow}. We exclude bot-only reviewed submissions, defined as agent-authored PRs without any human comments, to reduce cases where no human decision context is available. This step removes 1,249 PRs and yields a final dataset of 9,799 human-reviewed Agentic-PRs across multiple agents and repositories.

Table \ref{tab:agentcounts} summarizes the composition of the refined dataset by agent, reporting total, merged, and rejected PR counts. The dataset is unevenly distributed across agents: OpenAI Codex and Devin together account for the majority of Agentic-PRs, while Copilot and Cursor contribute smaller but substantial portions, and Claude Code appears least frequently. Across all agents, merged PRs are more common than rejected PRs, reflecting the overall composition of the refined dataset rather than agent-specific merge behavior.

\begin{table}[t]
\centering
\caption{Per-agent distribution of merged and rejected Agentic Pull Requests in the refined dataset after removing bot-only reviews (Figure~\ref{fig:flow}B).}
\label{tab:agentcounts}
\begin{tabular}{lrrrr}
\toprule
Agent & Total PRs & (\%) of Total & Merged & Rejected \\
\midrule
Claude Code    & 213   & 2.2  & 130   & 83   \\
Copilot        & 1{,}429 & 14.6 & 839   & 590  \\
Cursor         & 788   & 8.0  & 563   & 225  \\
Devin          & 3{,}380 & 34.5 & 1{,}813 & 1{,}567 \\
OpenAI Codex   & 3{,}989 & 40.7 & 2{,}834 & 1{,}155 \\
\midrule
Total          & 9{,}799 & 100.0 & 6{,}179 & 3{,}620 \\
\bottomrule
\end{tabular}
\end{table}

\section{Methodology}

Step C in Figure~\ref{fig:flow} depicts the stratified sampling strategy used for manual inspection. From the refined dataset, we draw a sample of 717 PRs, including 353 rejected and 364 merged submissions, balanced across agents and repositories. This stratification ensures representative coverage of decision outcomes and review interactions and follows prior recommendations for pull request–level empirical studies~\cite{wang2021benchmarkcodereviewstudies}.

The manual qualitative coding process is summarized in step D on Figure~\ref{fig:flow}. We conduct an exploratory empirical analysis of decision making in Agentic-PRs using manual qualitative coding, following established guidelines for empirical software engineering studies involving large language models~\cite{baltes2025guidelinesempiricalstudiessoftware}. The coding scheme is developed iteratively through pilot inspection and calibration. For rejected PRs (RQ1), cases are categorized as agentic failures, non-agentic failures, or unknown when no reliable decision signal is observable in comments, CI results, or closure context. For merged PRs (RQ2), interaction patterns are classified as feedback loops, human intervention, or no feedback loop, with an unknown label used when interaction evidence is missing or ambiguous.

Four annotators participate in the study, with two annotators independently coding each PR. Inter-rater reliability is assessed using Cohen’s $\kappa$, following best practices for agreement analysis in software engineering research~\cite{díaz2021applyinginterraterreliabilityagreement, krippendorff2011computing}. Disagreements are resolved through discussion, and ambiguous cases are conservatively assigned to the unknown category. All sampling scripts, coding guidelines, and agreement statistics are included in the replication package~\cite{agentic-prs-replication}.
\section{RQ1: What drives the rejection of Agentic-PRs?}

\begin{table}[t]
\centering
\caption{Distribution of rejection reasons ($N = 353$).}
\label{tab:rq1-rejection-types}
\begin{tabular}{lcccc}
\toprule
Agent & AF & Non-AF & Unknown & Total \\
\midrule
Claude Code    & 1  & 4  & 3  & 8   \\
Copilot        & 19 & 17 & 21 & 57  \\
Cursor         & 15 & 1  & 6  & 22  \\
Devin          & 37 & 69 & 47 & 153 \\
OpenAI Codex   & 54 & 19 & 40 & 113 \\
\midrule
Total          & 126 & 110 & 117 & 353 \\
Percentage     & 35.7\% & 31.2\% & 33.1\% & \\
\bottomrule
\end{tabular}

\vspace{2pt}
\footnotesize
\emph{Inter-rater reliability: Cohen’s $\kappa \approx 0.90$; AF = agentic failure; Non-AF = non-agentic failure.}
\end{table}

Table~\ref{tab:rq1-rejection-types} reports the distribution of rejection categories in the manually inspected sample of 353 rejected Agentic-PRs. Of these, 126 (35.7\%) were categorized as \emph{agentic failures} (AF), 110 (31.2\%) as \emph{non-agentic failures} (Non-AF), and 117 (33.1\%) as \emph{unknown}. AF cases were associated with observable technical failure signals such as failing checks, failing tests, or reviewer comments indicating that the submitted change did not work. Non-AF cases were associated with observable workflow- or process-related closure signals such as duplicates, superseded changes, inactivity, test PRs, or incorrect submission context. Unknown cases lacked sufficient observable evidence for classification and were often caused by silent repository behaviors (e.g., PRs closed without comments). These cases do not necessarily indicate agentic failure and should be treated separately or excluded from precision-based evaluations to avoid noise. Representative examples are provided in the replication package~\cite{agentic-prs-replication}.

The category composition varied across agents. For Devin, 37 of 153 rejected PRs (24.2\%) were categorized as AF, 69 (45.1\%) as Non-AF, and 47 (30.7\%) as Unknown. For OpenAI Codex, 54 of 113 rejected PRs (47.8\%) were categorized as AF, 19 (16.8\%) as Non-AF, and 40 (35.4\%) as Unknown. Copilot and Cursor showed distributions across all three categories, and Claude Code accounted for 8 rejected PRs in the sample, as summarized in Table~\ref{tab:rq1-rejection-types}.

\begin{center}
\fbox{
\begin{minipage}{0.92\linewidth}
\emph{RQ1 Answer.}
Rejection of Agentic-PRs was not primarily driven by agentic failures. Among 353 rejected Agentic-PRs, only 126 (35.7\%) were rejected due to observable agentic failures. The remaining rejections were split between non-agentic workflow-related causes (110 PRs, 31.2\%) and cases with no observable decision rationale (117 PRs, 33.1\%). Thus, a majority of rejected Agentic-PRs were not attributable to identifiable agentic failures.
\end{minipage}}
\end{center}

\section{RQ2: How does human involvement shape successful Agentic-PRs?}

\begin{table}[t]
\centering
\caption{Interaction patterns in merged PRs ($N = 364$).}
\label{tab:rq2-interactions}
\begin{tabular}{lccccc}
\toprule
Agent & FL & HI & No FL & Unknown & Total \\
\midrule
Claude Code   & 0  & 0  & 8   & 0  & 8   \\
Copilot       & 13 & 16 & 11  & 9  & 49  \\
Cursor        & 1  & 0  & 30  & 2  & 33  \\
Devin         & 14 & 11 & 74  & 8  & 107 \\
OpenAI Codex  & 0  & 1  & 165 & 1  & 167 \\
\midrule
Total         & 28 & 28 & 288 & 20 & 364 \\
Percentage    & 7.7\% & 7.7\% & 79.1\% & 5.5\% & 100.0\% \\
\bottomrule
\end{tabular}

\vspace{2pt}
\footnotesize
\emph{Inter-rater reliability: Cohen’s $\kappa = 1.0$; FL = reviewer feedback followed by agent-only revisions; HI = reviewer commits applied before merge.}
\end{table}

Table~\ref{tab:rq2-interactions} reports interaction patterns observed in 364 manually inspected merged Agentic-PRs. Most merged PRs, 288 (79.1\%), were categorized as \emph{No FL}, indicating merge without an observable feedback loop or reviewer-applied commits. In addition, 28 merged PRs (7.7\%) were categorized as \emph{FL}, and 28 (7.7\%) as \emph{HI}. The remaining 20 merged PRs (5.5\%) were categorized as \emph{Unknown} due to insufficient or ambiguous observable interaction traces.

The distribution of interaction patterns varied across agents. Copilot accounted for 29 of 49 merged PRs (59.2\%) with either \emph{FL} or \emph{HI} (13 FL; 16 HI), and Devin accounted for 25 of 107 (23.4\%) with either \emph{FL} or \emph{HI} (14 FL; 11 HI). OpenAI Codex showed 1 of 167 merged PRs (0.6\%) categorized as \emph{HI} and 0 categorized as \emph{FL}. Cursor showed 1 of 33 merged PRs (3.0\%) categorized as \emph{FL} and 0 categorized as \emph{HI}. Claude Code showed 0 cases categorized as \emph{FL} or \emph{HI} in the inspected sample. Representative labeled examples (including PR URLs and supporting interaction evidence) were provided in the replication package. \cite{agentic-prs-replication}

\begin{center}
\fbox{
\begin{minipage}{0.92\linewidth}
\emph{RQ2 Answer.}
Human involvement in successful Agentic-PRs was observed in 56 of 364 merged PRs (15.4\%), split evenly between feedback loops (28 PRs, 7.7\%) and reviewer-applied commits (28 PRs, 7.7\%). Most merged PRs (288 PRs, 79.1\%) were merged with no observed feedback loop, and 20 PRs (5.5\%) had ambiguous or insufficient interaction evidence. Human involvement was concentrated in Copilot and Devin PRs, which accounted for 54 of the 56 cases labeled as feedback loop or human intervention in the inspected sample.
\end{minipage}}
\end{center}

\section{Discussion}

Across the study, we find that \textbf{Agentic-PR outcomes cannot be reliably interpreted from merge or rejection labels alone}. Decisions were frequently communicated through interaction-level signals—such as CI behavior, review comments, commit history, and workflow actions—that were not consistently reflected in final PR states. In RQ1, 33.1\% of rejected PRs exhibited no observable comments or CI failures, and in RQ2 reviewer involvement was sometimes visible only through follow-up commits. Manual inspection shows that PRs may be closed after reviewers redirect effort, enforce repository conventions, or deprioritize changes without recording an explicit rejection rationale. For example, Copilot PR \#9653 in \texttt{dotnet/aspire} was closed without merge despite extensive discussion, multiple commits, and CI runs, illustrating how outcome-only analyses omit substantial decision context~\cite{golzadeh2019effect, cihan2024automatedcodereviewpractice}.

We also observe that \textbf{merging Agentic-PRs does not uniformly correspond to autonomous agent completion}. In RQ2, 15.4\% of merged PRs involved explicit reviewer participation, either through feedback followed by agent revisions or through reviewer-applied commits. These interventions commonly addressed missing tests, CI configuration issues, or repository-specific conventions rather than core functionality. A representative case is Copilot PR \#3512 in \texttt{primer/view\_components}, where a reviewer applied corrective commits before merging without further agent revision. Such cases show that merged outcomes may reflect collaborative completion rather than agent performance alone~\cite{sun2025doesaicodereview, palvannan2023suggestionbotanalyzingimpact}.

\textbf{Interaction patterns further varied systematically across agents and repository contexts.} Devin and Copilot accounted for most workflow-driven rejections and reviewer-involved merges, while OpenAI Codex and Cursor contributed a larger share of merges without revision. Qualitative inspection suggests these differences align with repository practices such as stricter CI gating, iterative review norms, and enforcement of contribution policies. Consequently, observed Agentic-PRs outcomes reflect both agent behavior and the review environments in which agents are deployed, complicating direct cross-agent comparisons.

\textbf{Limits of attribution motivate interaction-aware evaluation.}
Manual qualitative inspection reliably identified decision rationale when interaction evidence such as reviewer comments, CI outcomes, or corrective commits was present. However, silent closures and workflow-triggered actions lacked sufficient observable signals for attribution, revealing irreducible uncertainty in reconstructing decisions from PR artifacts alone. This suggests that both manual and automated evaluations should incorporate interaction evidence when available and explicitly represent uncertainty when it is absent, rather than inferring agent performance from outcome labels alone.

\section{Threats to Validity}

\textbf{Construct Validity.}
Our results depend on operational definitions of \emph{agentic failure}, \emph{non-agentic outcome}, and \emph{feedback loop}. Although these constructs were refined through pilot coding, ambiguity remains when reviewer intent is implicit or PR artifacts are sparse. Some decision rationale may be unobservable (e.g., silent closures), leading to an \emph{unknown} category that may hide unrecorded factors. 
We also lack explicit effort metrics (e.g., review duration, comment volume). Although 15.4\% of merges involved human intervention, the associated cost remains unquantified.

\textbf{Internal Validity.}
Manual coding may introduce subjectivity. We mitigated this through independent dual annotation and high inter-rater agreement (Cohen’s~$\kappa$), though subtle or multi-causal cases may still be misclassified. Stratified sampling reduces bias, but unobserved contextual factors cannot be fully controlled.

\textbf{External Validity.}
Our study focuses on high-star open-source repositories and five widely used coding agents. While this improves data quality, findings may not generalize to smaller projects, proprietary settings, or future agent systems, where workflows and review practices may differ.

\section{Conclusion and Future Work}

This study examined decision formation in Agentic Pull Requests by analyzing interaction artifacts from 11,048 closed PRs, including 717 manually inspected cases drawn from 9,799 human-reviewed submissions. We show that PR outcomes alone are unreliable indicators of agent performance: 33.1\% of rejected PRs lacked observable decision rationale, while only 35.7\% reflected clear agentic failures. Among merged PRs, 15.4\% required explicit reviewer involvement through feedback loops or direct human commits. We further observed systematic differences across agents, with Copilot and Devin more often embedded in workflow-heavy repositories and Codex and Cursor more frequently merged without interaction, indicating that observed outcomes reflect deployment context as much as agent-generated code.

These findings motivate evaluation approaches that move beyond outcome-based metrics and incorporate interaction-level evidence such as review comments, CI signals, commit histories, and authorship changes, while accounting for uncertainty when signals are absent. Future work should leverage manually curated ground truth to automate detection of workflow-driven closures, reviewer intervention, and feedback loops at scale, and develop interaction-aware benchmarks that better reflect agent effectiveness in real-world software development.

\bibliographystyle{ACM-Reference-Format}
\bibliography{reference}
\end{document}